\documentclass[a4paper,12pt]{article}
\usepackage{graphicx}
\usepackage{amsmath,amssymb,graphicx}
\usepackage{cite}
\usepackage{color}

\usepackage{slashed}
\numberwithin{equation}{section}

\newcommand{\ii}{\mathrm{i}}

\newcommand{\dd}{\mathrm{d}}

\newcommand{\M}{\mathcal{M}}

\newcommand{\e}{\mathrm{e}}

\newcommand{\tr}{\mathop{\mathrm{tr}}\nolimits}

\newcommand{\I}{\mathbb{I}}

\newcommand{\sign}{\mathop{\mathrm{sign}}}

\newcommand{\uU}{\mathrm{U}}

\begin{document}

\title{Spectral Dirac graphs}%
\author{Corneliu Sochichiu\thanks{e-mail:
\texttt{corneliu@gist.ac.kr}}\\
{\it GIST College \& Dept. of Physics and Photon Science,}\\
{\it Gwangju Institue of Science and Technology,}\\
{\it Gwangju, 61005, KOREA}
}%
%
%\subjclass{}%
%\keywords{}%

%\date{}%
%\dedicatory{}%
%\commby{}%
% ----------------------------------------------------------------
\maketitle
\begin{abstract}%:abstract
We address the problem of identifying families of discrete models naturally flowing  in continuum limit to relativistic quantum field theories. We call them Dirac graphs. In this work, we require the graphs to obey spectrality property, which implies that the adjacency matrix can be represented by a continuous function defined on the spectral space. We also consider deformations of such graphs away from the spectrality. We show that deformations with regular continuum limit result in background gauge field and gravity. Interestingly, gauge interactions appear due to contribution microscopic deformations while the standard gravity can only be a result of macroscopic (adiabatic) deformations.
\end{abstract}
%\maketitle
%\tableofcontents
% ----------------------------------------------------------------
\section{Introduction}\label{sec:Intro}%:Introduction
% --------------------------------------------------------------------
The picture of the Nature given by the modern physics includes the concept of continuous space-time which is a subject, at least locally, to Lorentz symmetry. The continuity of the space-time is an \emph{a priori} assumption, which is not based on any experiment (such an experiment would be not possible), at the same time, not contradicted so far by any observation.

Nevertheless, as pointed by R.~Penrose,  the concept of continuous space-time is faulty in diffeomorphism invariant theory of gravity, as in this case there is no bottom line for the count of degrees of freedom \cite{Penrose:1972jq}. To overcome this issue, he proposed the point of view in which the continuum space-time appears as an effective realm emerging from, as it can be put in modern language, entanglement of discrete degrees of freedom. String theory, on the other hand, predicts this would happen at the Hagedorn scale \cite{atick1988hagedorn}, when the space-time dissociates into individual ``points''.

Following above arguments, we have to admit that the continuity of the space-time should break at some scale and we have to deal with discrete models instead of particles and fields. Then, an interesting subject to study would be the properties of discrete models flowing to known field theories, and in particular, if there is any sort of universality in such behavior.

Hints for such models might come from condensed matter theory, where Lorentz symmetry emerges al low energies in some systems like e.g. graphene \cite{PhysRev.71.622} (see \cite{2009RvMP...81..109C} for a review). Quasiparticles in such systems `live' in an effective \( (2+1) \)-dimensional Minkowski space-time different from ours (e.g. where the speed of light is replaced by the Fermi speed). This effective Minkowski space is a result of correlations of electrons in the substance, and it is a fully macroscopic phenomenon.

Inspired by mentioned systems, it was researched if translational invariant lattices with similar properties exist in other dimensions as well \cite{Sochichiu:2011ap,Manes:2011jk}. It was found that Dirac lattices can be constructed in any number of dimensions when the lattice adjacency matrix satisfies a certain algebraic equation. Moreover, it appeared that in the case of even (space-time) dimensions the systems were stable against small translationally invariant perturbations when the number of dimensions is related to the cardinality of the lattice unit cell as \( D/2=2^{[d/2]} \), where \( D \) is the dimension of the translational invariance and \( [d/2] \) is the integer part of the half number of sites in the unit cell.\footnote{Generally, non-degeneracy of fermionic fluctuations requires  \( D/2 \leq 2^{[d/2]} \), the equality corresponding to the stable case.} This relation comes from the dimensionality of Clifford algebras associated to corresponding translation groups as described by ABS construction  \cite{Atiyah19643}. Also, the stability has a topological ground in the conservation of a K-theory charge \cite{PhysRevLett.95.016405}.
 
As we just pointed, in stable cases translational invariant deformations can not affect the continuum limit. Translation non-invariant deformations instead can lead to emergence of background gauge and gravity fields when they have a well-defined continuum limit \cite{Sochichiu:2010ns}.\footnote{In \( (2+1)\)-dimensional case, however,  the emergent gravity background can be absorbed into redefinition of the gauge field.} 

Since the time parameter is pre-existent in Dirac lattice, the problem is that deformations can control only the spacial part of metric or gauge potential. Hence, there is no possibility to produce a generic gauge or gravity configuration. Therefore, we considered a further step in generalization to discrete structures were also the time extension becomes effective. We call such structures \emph{Dirac texture graphs}. Such graphs with four-dimensional translational symmetry were studied in \cite{Sochichiu:2014gha}. The non-degeneracy and stability in the case of four-dimensional translational symmetry require that the unit cell of such graphs consist of two sites. Since this is the simplest unit cell after one consisting of a single site which leads to \( (1+1) \)-dimensional theory like in Tomonaga-Luttinger model (see \cite{0022-3719-14-19-010} for a review), we will restrict ourselves throughout this work to the \emph{two-site unit cell}.

Translational symmetry in the above models was important as a technical limitation in achieving the continuum limit, as it allowed to use the tools of Fourier transform. In this work we relax this limitation and consider \emph{spectral graphs} instead, which is a more general discrete structure which includes translation invariant graphs, group lattices, and others as particular cases. We define these graphs below in terms of spectral images of their adjacency matrices. For our two-site unit cell case, these are generic continuous \( 2 \times 2 \)-matrix functions on a spectral space. We choose spectral space to be a smooth manifold, but various generalizations to more singular structures are also possible. In such an approach the concept of \emph{locality} becomes also effective: now it can be defined only macroscopically, in terms of the continuous geometry. We consider this not to be a defect of our approach, but rather an interesting feature.

The plan of the paper is as follows. In the next section, we define the spectral graph system which is the starting point for our analysis.  Then,  in sec.\ref{sec:cont-lim} we describe the continuum limit of such graphs. In sec.\ref{sec:def} we proceed to the analysis of deformations in the continuum limit and identify those deformations which are leading to gauge and gravity backgrounds. Finally, we discuss the results. Appendix section includes some properties of Clifford algebra structures  used in the main part of the work.
 
% --------------------------------------------------------------------
\section{The model: Spectral graphs}\label{sec:Model}%:Section

Let us consider a network of entangled fermionic states, \( \theta_{n} \) with the gaussian entanglement on a graph. The state of the system is given by the microscopic density matrix \( \rho \) given by,
\begin{equation}
 	\rho \propto \e^{\ii \theta^{\dag} \cdot T \cdot \theta},
\end{equation}
where the entanglement is encoded in the matrix \( T \) with elements \( T_{mn} \) which is the graph's  adjacency matrix. 

Relevant for a macroscopic observer, the reduced density matrix  is given by integration over microscopic fields \( \theta_{n} \). The reduced state is described by the following partition function,
\begin{equation}\label{model:PI}
 	Z(T)= \int [\dd \theta^{\dag} \dd \theta] \e^{\ii S_{0}[ \theta^{\dag}, \theta]}, \qquad
	S_{0}[ \theta^{\dag}, \theta]=\theta^{\dag} \cdot T \cdot \theta.
\end{equation}
The integration is assumed to be over all fermionic degrees of freedom apart from those  accessible to the continuum observer. The meaning of this must become clear later.

In \cite{Sochichiu:2014gha} it was translational invariance of the adjacency matrix which was used to define continuum limit of the lattice. Translational invariance allows introduction of the Fourier transform. Then, the continuum limit can be defined in terms of the Fourier momentum. Now instead of the translational invariance we will use the \emph{spectrality} property, which will be defined below. Also in this work we we do not require any form of graph locality, like nearest neighbour nature or so.  The continuum space description will be achieved from the spectrum of the fermionic adjacency matrix \( T \). Hence, the locality at the macroscopic level will emerge as a property of the adjacency matrix.

Let us consider following structures: A unit cell consists of \( d_{0} \) nodes, which we call dimension of unit cell.  In this work we will use the minimal non-trivial case of \( d_{0}=2 \). Unit cells in the graph are labelled by index \( n \). The fermionic field on the graph is represented by a \( d_{0} \)-component Grassmann function \( \theta_{n} \). 

A \emph{spectral graph} by definition is such a graph for which graph fields exist as a spectral transform, a generalisation of Fourier transforms, e.g. any field \( \theta_{n} \) can be represented by its spectral image  \( \theta(k) \), as follows,
\begin{equation}\label{Four-tr}
 	\theta_{n}= \int_{\M_{k}} \dd k R_{n}(k) \theta(k),
\end{equation}
where the integration is over the spectral space \( \M_{k} \) endowed with the measure \( \dd k \), while \( R_{n}(k) \) is the spectral representation function. 

We will tacitly assume \( \M_{k} \) to be a \( D \)-dimensional manifold, although smooth manifold properties are not strictly necessary. 

In the case of our model, the requirement of non-degeneracy of fermionic fluctuations near the saddle points of the action \( S_{0} \) will impose certain  constraints on \( \M_{k} \) for specific \( d_{0} \). 

An important property required from the spectral representation function is the invertibility of the spectral transform, which implies the existence of the conjugate function \( R_{n}^{*}(k) \) such that,
\begin{equation}
 	\int_{\M_{k}} \dd k R^{*}_{n}(k)R_{m}(k)= \delta_{nm}, \qquad
	\sum_{n} R^{*}_{n}(k)R_{n}(k')= \delta(k,k'),
\end{equation} 
where \( \delta_{nm} \) the graph's Kronecker symbol, while \( \delta(k,k') \) the delta function on \( \M_{k} \). 

Finally, we require that the inverse Fourier transform to \eqref{Four-tr} diagonalises the adjacency matrix, which means that the adjacency matrix can be represented as,
\begin{equation}
 	T_{nm}= \int_{\M_{k}} \dd k R_{n}(k) T(k) R^{*}_{m}(k),
\end{equation}
where \( T(k) \) is a \( d_{0} \times d_{0} \) matrix function. 

In other words a graph is spectral if the eigenvalues of its adjacency matrix can be organized into those a \( d_{0} \times d_{0} \) matrix function over a continuum space endowed with some integration measure.

As an example, the uniform lattice is given by the spectral representation function \( R_{n}(k)=\e^{\ii k \cdot n} \) corresponding to the ordinary Fourier transform.

As we mentioned already, K-theory considerations \cite{Atiyah19643} establish a relation between the dimensionality of the unit cell \( d_{0} \) and the dimension \( D \) of the spectral space. As it appears \cite{Sochichiu:2011ap}, non-degeneracy of fermion fluctuations near zeroes of \( T \) requires that \( D \leq 2^{[d_{0}/2]+1}\), while the stability against translational invariant deformations requires the condition to saturate into the equality: \( D=2^{[d_{0}/2]+1} \) \cite{PhysRevLett.95.016405}. 

According to the above arguments, the case of \( d_{0}=1 \) corresponds to two-dimensional models, as in Tomonaga-Luttinger fermionic model \cite{Luttinger:1963zz,Tomonaga:1950zz}.  The next simplest case is \( d_{0}=2 \), stabilised by \( D=4 \). This case brings to four-dimensional quantum field models. In what follows we limit ourself to this case, i.e. from now on we have, \( d_{0}=2 \) and \( D=4 \).

% --------------------------------------------------------------------
\section{Continuum limit}\label{sec:cont-lim}
% --------------------------------------------------------------------
Assuming the spectrality property of the adjacency matrix, as well as setting \( D=4 \) and \( d_{0}=2 \) we can express the the action in the following form,\footnote{For later convenience we normalise the measure to contain the \( (2\pi)^{-4} \) factors.}
\begin{equation}\label{model:action}
	S_{0}=\int_{\M_{k}} \frac{\dd^{4} k }{(2\pi)^{4}} \theta^{\dag}(k) \cdot T(k) \cdot  \theta(k).
\end{equation}

The continuum limit is determined by the leading contribution in the saddle point approximation, which is given by zeroes of the adjacency matrix \( T \). In terms of the spectral function \( T(k) \), zero modes correspond to points in the spectral space with vanishing determinant of  \(  T(k) \). 

Expanding the adjacency matrix \( T(k) \) in terms of extended set of Pauli matrices  \( \sigma_{A}\), \( A=0,1,2,3 \),
\begin{equation}\label{T-expn}
 	T(k)= T_{A}(k) \sigma^{A},
\end{equation}
the zero determinant condition takes the form,
\begin{equation}\label{T-det-ind}
 	\det T \equiv T_{0}^{2}-T_{i}^{2}=0,
\end{equation}
where we assume sum over \( i=1,2,3 \).

The indefinite signature in eq. \eqref{T-det-ind} makes the continuum limit ill-defined. Following \cite{Sochichiu:2014gha}, we circumvent this problem by a Wick rotation of \( T_{0} \) component of the adjacency matrix,
\begin{equation}
 	T_{0}\to \ii T_{0}, \qquad k_{0}\to \ii k_{0}.
\end{equation}
Then the original partition function is defined as an analytic continuation of the Wick rotated one. From now on we assume that this is the way how the partition function is defined.

In the `Wick rotated' picture vanishing of the determinant requires vanishing of each function \( T_{A}(k) \). Therefore the low energy contribution comes from regions near points  \( K_{ \alpha} \), \( \alpha=1, \dots, \nu \), such that,
\begin{equation}
 	T_{A}(K_{ \alpha})=0.
\end{equation}
These points we call \emph{pseudo-Fermi points} (pFp).

We consider the situation when pFp are isolated zeroes.  In contrast to \cite{Sochichiu:2014gha}, we consider pFp's of arbitrary finite degrees, not just linearly non-degenerate ones. 

At each pFp we can introduce local `coordinates' \( p_{A} \approx T_{A}\) covering its vicinity. Unless the pFp is of first degree these coordinates \( p_{A} \) are multivalued rather than covering the vicinity one-to-one.
Indeed, for a pFp \( K_{ \alpha} \) of degree \( n_{ \alpha} \) the equation,
\begin{equation}
 	p_{A}= T_{A}(k),
\end{equation}
algebraically has \( n_{ \alpha} \) solutions near the point \( K_{ \alpha} \),  \( n_{ \alpha} \) being also the topological index of the map \( T_{A}(k) \). This is not a problem: We can count solutions by labelling the sheets by an index \( a_{ \alpha}=1, \dots , n_{ \alpha} \). This is corresponding to one pFp, \( K_{ \alpha} \). When we want to count all solutions for all pFps we will use the index \( a \), i.e. \( a \) runs through all \( \sum_{ \alpha} n_{ \alpha} \) values. 

As a result of transformation the action \eqref{model:action} localises near zero points where with all above notations it takes the form,
\begin{multline}\label{ferm-act}
 	S_{0} \approx 
	\sum_{ \alpha}\int_{k\sim K_{ \alpha}} \frac{\dd^{4}k}{(2\pi)^{4}}  \theta^{\dag} (k) 
	(T_{A}(k) \sigma^{A}) \theta(k)\\
	=
	\sum_{a} \int \frac{\dd^{4}p }{(2\pi)^{4}} h_{(a)}^{-1}
	\theta^{\dag}_{a} (p) 
	(p_{A} \sigma^{A}) \theta_{a}(p)
	, 
\end{multline}
where \( h_{(a)}^{-1} \) is the value of the transformation Jacobian on the \( a \)-th sheet of the map \( p \to k \), i.e.,
\begin{equation}
 	h_{(a)}=\left.\det \left( \frac{ \partial T}{ \partial k}\right)\right|_{k=k_{a}(p)}.
\end{equation}

Let us note, that due to topological reasons, the total degree of zeroes of \( T(k) \), taking into account the sign of the determinant \( h_{( \alpha)} \) should vanish for an orientable \( \mathcal{M}_{k} \),
\begin{equation}\label{cond:zero-chirality}
 	\sum_{ \alpha} \sign h_{( \alpha)} n_{ (\alpha)}=0.
\end{equation}
This implies that the cumulative degree \( n_{+}= \sum_{h_{( \alpha)}>0} n_{( \alpha)} \) of pFps with positive \( h_{( \alpha)} \) must be equal to the cumulative degree \( n_{-} \) of points with  \( h_{( \alpha)}<0 \), i.e. there are in total an even number of modes \( \theta_{a} \), half of which emerging at points \( K_{ \alpha} \) with the positive determinant \( h_{(\alpha)} \), and half emerging at points with the negative \( h_{( \alpha)} \). 

We can combine the positive modes \( \theta_{a} \) into the positive chirality multiplet \( \psi_{+} \) according to,
\begin{subequations}\label{psi:def}
\begin{equation}
 	\psi_{+,a}(p)= |h_{ (a)}|^{-1/2} \theta_{a}(p), \qquad a=1,\dots, n_{+},
\end{equation}
and, respectively, the negative modes into the negative chirality multiplet,
\begin{equation}
 	\psi_{-,\dot{a}}= |h_{ (a)}|^{-1/2}  \sigma^{2}(\theta^{\dag})^{T}_{a}(p), \qquad a=1,\dots, n_{-},
\end{equation}
\end{subequations}
where \( {T} \) stands for the transposition of the cell index, and \( \sigma^{2} \) is the Pauli matrix acting on the cell index. 

After all identifications, we can re-write the action \eqref{model:action} in the following form,
\begin{equation}\label{almost-cont}
 	S_{0}= 
	\int \frac{\dd^{4}p}{(2\pi)^{2}} \bar{ \psi}(p) \gamma^{ A} p_{ A} \psi(p),
\end{equation}
where we introduced the Dirac conjugate \( \bar{ \psi}= \psi^{\dag} \gamma^{0} \), as well as defined the set of Dirac matrices (see the Appendix),
\begin{equation}\label{DiracMat}
 	\gamma^{0}=\I \otimes \sigma^{1}, \qquad
	\gamma^{a}= \gamma^{0} \cdot (\sigma^{a} \otimes \sigma^{3})=
	-\ii (\sigma^{a} \otimes \sigma^{2}).
\end{equation}

Applying  the  inverse continuous Fourier transform,
\begin{equation}\label{psi-inv-four}
 	\psi (x)=
	\int \frac{\dd^{4}p}{(2\pi)^{4}}
	\e^{- \ii p \cdot x} \psi(p),
\end{equation}
to the action \eqref{almost-cont}, we can express the result in the standard form of Dirac action for  \( n=n_{+}=n_{-} \) positive and the same number of negative chiral fermions,
\begin{equation}\label{cont-act0}
 	S_{0}=
	-\ii\int \dd x \bar{ \psi}  \slashed\partial \psi,
\end{equation}
where \( \slashed\partial = \gamma^{ \mu} \stackrel{\leftrightarrow}{\partial}_{ \mu}\), is the (symmetric) Dirac operator.

So far, we obtained a collection of free Weyl fermions of both chiralities, which, in principle, can be combined into \( n \) Dirac fermions with \( \uU(n) \) vector and \( \uU(n) \) axial global symmetries, but now let us turn to deformations of the graph.

% --------------------------------------------------------------------
\section{Deformations}\label{sec:def}
Let us consider a deformation of the adjacency matrix \( D=T+G \), where \( G \) is the deformation part. It is given by the spectral kernel \( G(k,l) \), such that in the most general case the new deformed action takes the form,
\begin{equation}
	S \equiv S_{0} + \Delta S,	
\end{equation}
where the deformation is given by,
\begin{equation}
 	\Delta S=
	\int \frac{\dd^{4}k}{(2\pi)^{4}} \frac{ \dd^{4}l}{(2\pi)^{4}}
	 \theta^{\dag}(k) G(k,l) \theta(l),
\end{equation}
where the kernel \( G(k,l) \) is \( 2 \times 2 \) matrices in unit cell index.

We assume the deformation \( G \) small in the sense, that the partition function for the deformed graph is still dominated by regions near zeroes of the undeformed \( T \). Also, it is important that the deformation has a well-defined continuum limit in terms of local fields. Therefore, we shall restrict ourselves to deformations which are localised both on the graph and spectrally. 

The spectral locality can be implemented by choosing  deformations with the kernel \( G(k,l) \), which is vanishing quickly with separation of \( k \) and \( l \) in the spectral space, while the graph locality would imply that the deformation vanishes with the distance on the graph. The problem with this property is that the distance, as we define it both on the graph and in the spectral space, depends on the adjacency matrix itself. However, in the case of small deformations we can treat the problem perturbatively, i.e. use definition of locality with respect to undeformed adjacency matrix. At least, we have to assume that the range of the deformation is less than the distance between distinct pFps, otherwise we will end up with terms breaking Lorentz invariance. This situation is very different from the 2D case, where long range deformations actually produce non-abelian gauge fields \cite{Sochichiu:2010ns}.

As the main contribution still comes from the modes in the vicinity of zeroes of \( T \), the deformation action restricted to such modes takes the form,
\begin{equation}\label{def-act}
 	\Delta S=
	\sum_{a,b}\int \frac{\dd^{4}p}{(2\pi)^{4}} \frac{\dd^{4}q}{(2\pi)^{4}}
	h_{a}^{-1} h_{b}^{-1}\theta^{\dag}_{a}(p) \tilde{G}_{a,b}(p,q) \theta_{b}(q),  
\end{equation}
where the components of \( \tilde{G}_{a,b} \)  are given by by branches of \( G(k,l) \) in coordinates \( p \)-\( q \),
\begin{equation}\label{tildeG}
 	\tilde{G}_{a,b}(p,q)=G(k_{a}(p),k_{b}(q)) .
\end{equation}

As a \( 2 \times 2 \) unit cell matrix, the kernel can be expanded in terms of the extended set of Pauli matrices,
\begin{equation}
 	\tilde{G}_{a,b}(p,q)=\tilde{G}^{A}_{a,b}(p,q) \sigma_{A},	
\end{equation}
such that the deformation part \eqref{def-act} of the action takes the form,
\begin{equation}\label{def-act-sigma}
 	\Delta S=
	\sum_{a,b}\int \frac{\dd^{4}p}{(2\pi)^{4}} \frac{\dd^{4}q}{(2\pi)^{4}} \tilde{G}^{A}_{a,b}(p,q)
	h_{a}^{-1} h_{b}^{-1}\theta^{\dag}_{a}(p)  \sigma_{A}\theta_{b}(q).
\end{equation}

The deformation \eqref{def-act-sigma} is non-local, but let us assume that at least in the vicinity of pFps the near-local expansion of the the deformation field \( \tilde{G}^{A}_{a,b}(p,q) \) holds,
\begin{equation}\label{def-field}
 	\frac{\tilde{G}^{A}_{a,b}(p,q)}{|h_{a}(p)|^{1/2} |h_{b}(q)|^{1/2}}= g^{A}_{a,b}(p-q)+
	\frac{1}{2} g^{AB}_{a,b}(p-q) (p+q)_{B}+ \dots,
\end{equation}
with some functions \( g^{A}_{ab} \), \( g^{AB}_{ab} \), etc, depending only on the difference of momenta. Dots stay for higher powers in \( (p+q) \). In terms of this expansion the deformed action takes the following form when restricted to the leading two terms,
\begin{multline}\label{def-act-exp}
 	\Delta S=
	\sum_{a,b}\int \frac{\dd^{4}p}{(2\pi)^{4}} \frac{\dd^{4}q}{(2\pi)^{4}} \times \\
	\left(\frac{1}{2}  \psi^{\dag}_{a}(p)g^{AB}_{a,b}(p-q) (p+q)_{B}\sigma_{A} \psi_{b}(q)
	+ \psi^{\dag}_{a}(p) g^{A}_{ab}(p-q) \eta_{b}(q)\right).
\end{multline}

Performing the continuum space inverse Fourier transform on \eqref{def-act-exp}, the deformation action becomes as follows,
\begin{equation}\label{def-act-cont}
 	\Delta S= \int \dd x \sum_{ \alpha}\bar{ \psi}_{ \alpha} 
	\left\{
	\gamma^{A} g^{ B}_{(\alpha)A} (-\ii \partial_{ B})+ g_{(\alpha)A} \gamma^{A} 
	\right\}
	\psi_{ \alpha},
\end{equation}
where we explicitly separated the modes coming from a separate pFp \( K_{ \alpha} \) into a chiral field \( \psi_{ \alpha} \). Chiral fields \( g_{(\alpha)A} \) and \( g^{B}_{( \alpha)A} \) belong to adjoint representations of \( \mathrm{U}(n_{ \alpha}) \) associated to  \( K_{ \alpha}\).
 
As it appears from \eqref{def-act-cont}, the resulting deformed action describes a system of gauge non-abelian fermionic multiplets \emph{generally} living on separate space-time sheets. In other words, the space-time geometry according to the deformed action can be flavour dependent. The flavour dependence is a  violation of the universality property of gravity, or of the \emph{general equivalence principle}. 

A `healthy' approach would be to restrict the class of deformations in order to `promote' only physically sound degrees of freedom. We will do this in a while. Before that we can wonder whether the flavour dependent gravity can be given any physical reason. Up to our knowledge, the general equivalence principle of gravity was  experimentally checked only for \emph{neutral} matter. 

In what follows we propose two different classes of graph deformations leading to a meaningful continuum theory. We call these limits \emph{macroscopic (adiabatic)} and \emph{ultra-local} deformations, respectively. The two types combine into the coupling to the external gauge field and space-time geometry in the form of spin connection. Let us consider them one-by-one.

\paragraph{Ultra-local deformations.} 
The local modes are the modes which are independent or depend weakly on the sum momenta \( (p+q) \). On the other hand, their spectral range can extend to the whole neighbourhood of a Fermi point. Thus, if we introduce a UV cutoff scale \( \Lambda \) such that the continuum model is properly approximating the graph when momenta are within the range \( \Lambda \) from a Fermi point, i.e. \( p \lesssim \Lambda\), then the cutoff \( \Lambda \) will be effectively also the range of ultra-local deformations. 

In this case the resulting contribution to the continuum action becomes limited to the second term in the eq. \eqref{def-act-cont},
\begin{equation}\label{def-ultra-loc}
 	\Delta S_{\text{ultra local}}= \sum_{ \alpha}
	\int \dd x \bar{ \psi} g_{( \alpha)A} \gamma^{A} \psi,
\end{equation}
with the chiral gauge field \( g_{( \alpha)A}(x) \).

The gauge group is  the product over Fermi points of factors,
\begin{equation}\label{gauge-group}
 	G= \prod_{ \alpha} \mathrm{U}(n_{ \alpha})_{ \epsilon_{ \alpha}},
\end{equation}
where \( \epsilon_{ \alpha}= \sign h_{( \alpha)}\) is the chirality of the Fermi point \( \alpha \). 

Although, apparently chiral, the finiteness of the theory together with explicit gauge invariance implies that the theory is anomaly free. In particular the equation \eqref{cond:zero-chirality} is equivalent to the necessary condition for the chiral gauge anomaly cancellation (see e.g. \cite{Faddeev:1980be}). 

\paragraph{Macroscopic (adiabatic) deformations.}
Let us consider deformations localised in the spectral space. That is, we look at deformation functions \( \tilde{G}(k,l) \) essentially vanishing for arguments away from pFp by distance of order \( \mu_{g} \ll \Lambda \). The spectral space distance can be defined perturbatively in terms of induced metric,
\begin{equation}
 	\dd s_{k}^{2}\equiv g^{ij}\dd k_{i} \dd k_{j} = \frac{1}{2}\tr \dd T \dd T.
\end{equation}

This restriction results in the diagonal form of deformation fields. In principle, there could be a dependence on the pFp, but we further have to restrict deformations to be independent of pFp. The fact that the deformation does not distinguish between pFps implies that the scale of the deformation is macroscopic. 

The resulting deformation fields  in eq. \eqref{def-field} appear as follows,
\begin{equation}\label{gra-def-prop}
 	g_{Aab}(p-q)= \delta_{ab} g_{A}(p-q), 
	\qquad
	g_{Aab}^{B}(p-q)= \delta_{ab} g_{A}^{B}(p-q),
\end{equation}
and the corresponding deformation of the continuum action takes the form,
\begin{equation}\label{def-long-wave}
 	\Delta S_{\text{s.l.}}=
	\int\dd x \left(\frac{-\ii }{2}\bar{ \psi}  \gamma^{A} ( g_{A}^{B} \partial_{B} + \partial_{B} g_{A}^{B})\psi
	+ \bar{ \psi} g_{A} \gamma^{A} \psi
	\right),
\end{equation}
where Abelian fields \( g_{A}^{B}(x) \) and \( g_{A}(x) \) are the continuum Fourier transform of respective deformation fields \( g_{A}^{B}(p) \) and \( g_{A}(p) \).  

When added to the bare continuum action  \eqref{cont-act0} the above deformation \eqref{def-long-wave} produces a kinetic term resembling in the linear approximation the action for a Dirac fermion in a curved background geometry, 
\begin{equation}\label{bare-and-sl}
 	S_{0}+\Delta S_{\text{s.l.}}=
	\int\dd x \sqrt{-g} \left(\frac{-\ii }{2}\bar{ \psi}  \gamma^{A} ( \xi_{A}^{ \mu} \partial_{ \mu} + \partial_{ \mu} \xi_{A}^{ \mu})\psi
	+ \bar{ \psi} g_{A} \gamma^{A} \psi\right),
\end{equation}
where \( \xi^{ \mu}_{A} \) defined as \( \xi_{A}^{ \mu}+ \delta^{ \mu}_{A} \xi= \delta_{A}^{ \mu}+ g_{A}^{ \mu} \) is the vierbein field, which is a linear perturbation from the flat Minkowski background, here \( \xi= \xi^{A}_{A} \). This action, however, fails to reproduce a generic spin connection at once.

Comparing the action \eqref{bare-and-sl} to the irreducible Clifford algebra basis decomposition \eqref{cov-fer-mat} of a generic covariant derivative in the Appendix, we can notice that the fermionic operator fails to reproduce several terms of the generic spin connection. Thus, third term (and, optionally, the fourth one if one wants to include torsion)  of eq. \eqref{cov-fer-mat} is not reproduced in the expansion of \( \gamma^{0} \gamma^{ \mu}D_{ \mu}^{\text{s.l}} \), where,
\begin{equation}\label{spin-conn-trunk}
 	D^{\text{s.l.}}_{ \mu}=
	\partial_{ \mu}+ \ii g_{ \mu},
\end{equation}
and \( \gamma^{ \mu}= \xi_{A}^{ \mu} \gamma^{A} \). The spin connection, as it appears in \eqref{spin-conn-trunk} is truncated to just its `vector' part. This still does not create a problem, because the missing part of the generic spin connection can be recovered from the Abelian part of the chiral gauge field emerging in the ultra-local deformation, or alternatively by a redefinition of the fermionic field.

Summarising our attempt to single out the deformations leading to background geometry coupling, we were able, at the best, to define the gravity background by restricting the deformations of the adjacency matrix to something that looks like adiabatic form, which is in essence macroscopic. This suggest the macroscopic and effective nature of gravity emerging in our picture.

To conclude, let us express the fermionic action including combined contributions \eqref{def-ultra-loc} and \eqref{def-long-wave} in the standard form,
\begin{equation}
 	S_{0}+ \Delta S_{\text{u.l.}}+ \Delta S_{\text{s.l.}}=
	\int \dd x \sqrt{-g} \bar{ \psi} \gamma^{ \mu} D_{ \mu} \psi,
\end{equation}
where the covariant derivative takes the from,
\begin{equation}
 	D_{ \mu}=
	\partial_{ \mu}+ \omega_{ \mu AB} \gamma^{AB}+g'_{(\alpha) \mu},
\end{equation}
where \(  \omega_{ \mu AB} \) is the standard spin connection constructed from the vierbein field \( \xi_{A}^{ \mu} \), and \( g'_{(\alpha) \mu} \) is the modified gauge field after shifting by spin connection terms.

We expect that the fermionic back reaction will generate invariant dynamical terms for the gauge field and gravity background, which in the leading approximation must take the form of Yang-Mills and Einstein-Hilbert actions \cite{Sakharov:1967pk}.

Let us note, that in the case of only ultra-local deformations there is a stability of the continuum theory in the following sense. Inclusion of the Yang-Mills dynamical part for the gauge field should lead to a renormalisable theory.  Then, small graph deformations moving away from ultra-locality are given by  operators which are IR irrelevant. This means that they scale down in the continuum limit, leaving only the ultra-local contribution. 

% --------------------------------------------------------------------
\section{Discussions}
In this work, we considered spectral Dirac graphs as well as their deformations. Spectral graphs are a generalization of translation invariant graphs considered in earlier works \cite{Sochichiu:2011ap,Sochichiu:2014gha}. Spectrality does not imply any local structure, like the nearest-neighbor or etc. character of the graph.

The continuum limit in our model is defined by pseudo-Fermi points which are finite degree zeroes of the spectral image of the adjacency matrix. The fermionic degrees of freedom around such points are generically given by Weyl fermions. Higher degrees zeroes lead multiplets of Weyl fermions with the global symmetry group \( \mathrm{U}(n) \), where \( n \) is the degree of zero. 

We considered the setup in which the unit cell consists of two sites, which leads to a four-dimensional continuum model stable against small deformations, as predicted by K-theory arguments \cite{PhysRevLett.95.016405}. Although, every single pseudo-Fermi point produces a  multiplet of \emph{chiral} fermions, in the case of compact orientable spectral space, topological arguments similar to used for Nielsen-Ninomiya theorem in lattice theory guarantees that the total number  of fermionic degrees of freedom of both chiralities match.
  
Furthermore, we consider deformations of spectral graphs with a well-defined continuum limit and are `close' to local in the sense of background geometry.  Among such deformations, we distinguish two types. One type, which we call \emph{ultra-local deformations} produce the gauge field background coupled to the fermionic particle. Upon inclusion of the fermionic back reaction, the gauge field is expected to acquire a renormalizable gauge invariant dynamical part. Because this deformation is given by marginal operators, we expect the class of these deformations to be stable in the continuum limit, when the difference is given by irrelevant operators.

The other type of deformation we call \emph{macroscopic or adiabatic}. These deformations, which can be `naturally' selected from the graph structure lead  in the continuum limit to a coupling of the fermionic field to the space-time geometric background in the form of a vierbein field and a part of the spin connection. The missing part can be retrieved either from the ultra-local contribution or by a local redefinition of the fermionic field.

In the case we consider the \emph{full spectrum} of deformations with well-defined continuum limit, we have to face the situation of flavour-dependent gravity background. Although, the existence and flavour dependence of the microscopic gravity is not ruled out experimentally, and probably can not be checked at the present time, because of weak nature of gravity at the subatomic level, we expect that more careful continuum limit including the fermionic back reaction will suppress this interaction at relatively low energies, by the same RG reason we discussed in the gauge interaction case. The fermionic back reaction generated dynamical terms for the gravity (flavour dependent or not) result in a \emph{non-renormalizable theory}. According to a standard quantum field theory argument \cite{weinberg1995quantum} the non-renormalizable interaction eventually dies off in the continuum limit, so only infrared part of the deformations can contribute there.

Unfortunately, dynamical parts of gravity and gauge fields as well as the interaction couplings depend strongly on the details of the spectral space. We leave for a future work the task to find such a dependence on geometrical properties of the spectral space.
% --------------------------------------------------------------------
\appendix
% --------------------------------------------------------------------
\section{Dirac matrices, spin connections and other structures}\label{appnd}

In the main body of paper we see fermionic modes combining into familiar Dirac structures. Dirac action depends explicitly on combinations of Dirac matrices like \( \gamma^{0} \gamma^{a} \), for the spacial index \( a=1,2,3 \), found to be,
\begin{equation}\label{gamma0a}
 	\gamma^{0} \gamma^{a}= \sigma^{a} \otimes \sigma^{3},
\end{equation}
where the first factor in the tensor product is the unit cell space, while the second factor is the chirality of the Fermi point. The matrices \eqref{gamma0a} are compatible with the following choice of \( \gamma^{0} \),\footnote{Found by trial-and-error.}
\begin{equation}
 	\gamma^{0}= \I \otimes \sigma^{1},
\end{equation}
which leads to \eqref{DiracMat}.

In the gravity background the Dirac particle is coupled through the covariant derivative,
\begin{equation}
 	 D_{ \mu}= \partial_{ \mu}+ \frac{1}{4} \omega_{ \mu}{}^{AB} \gamma_{AB}+V_{ \mu}+A_{ \mu} \gamma_{5},
\end{equation}
where \( \partial_{ \mu} \) is the coordinate derivative, \( \omega_{ \mu}{}^{AB}  \) are the coefficients of spin connection, and \( \gamma_{AB} = \frac{1}{2}[ \gamma_{A}, \gamma_{B}]\), the antisymmetric product of coordinate-independent Dirac gamma matrices of an orthonormal frame. Decomposition of the matrix \( \gamma^{0} \gamma^{ \mu} D_{ \mu} \) appearing in the fermionic action in terms of irreducible generators of the four-dimensional Clifford algebra gives,
\begin{multline}\label{cov-fer-mat}
 	\gamma^{0} \gamma^{ \mu} D_{ \mu}=
	\left( h_{0}^{ \mu} \partial_{ \mu}+V_{0}+ \frac{1}{4} \omega_{a}{}^{[0a]}\right) \I \\
	+\left( h_{a}^{ \mu} \partial_{ \mu}+V_{a}+ \frac{1}{4}\omega_{0}{}^{[0a]}+
	\frac{1}{2}\omega_{c}{}^{[ca]}\right) \gamma^{0a} \\
	+\left( \frac{1}{4}\omega_{0ab}- \frac{1}{4} \omega_{a[0b]}+ \epsilon_{abc}A^{c}\right) \gamma^{ab}\\
	+\left(A_{0}+ \frac{1}{4} \epsilon_{abc}\omega^{abc}\right) \gamma^{5}.
\end{multline}

In terms of our representation, the Clifford algebra generators appearing in \eqref{cov-fer-mat} are as follows,
\begin{align}
	\I & = \I \otimes \I \\
	\gamma^{0a}&= \sigma_{a} \otimes \sigma_{3} \\
 	\gamma^{ab} &
	=-\ii \epsilon^{abc} \sigma_{c} \otimes \I,\\
	 \gamma^{5}&=  
	 \I \otimes \sigma^{3}.
\end{align}

We can compare term-by-term the fermionic operator appearing in the generic deformed action \eqref{def-act-exp},
\begin{multline}
 	\Delta D=
	\left( \frac{1}{2} \eta_{0}^{B} \partial_{B}+ a_{0}\right) \I 
	+\left( \xi^{aB} \partial_{B} +v^{a}\right) \gamma^{0a}
	+\left( \frac{1}{2} \eta_{a}^{B} \partial_{B}+ a_{a}\right) \epsilon_{abc} \gamma^{bc} \\
	+\left( \frac{1}{2} \xi_{0}^{B} \partial_{B}+ v_{0} \right) \gamma^{5}.
\end{multline}

% --------------------------------------------------------------------
\bibliographystyle{hunsrt}
\bibliography{nonfermi1}%:select bib file
	
\end{document}